\documentclass[twocolumn,pr,notitlepage,longbibliography]{revtex4-1}
\usepackage{amsmath}
\usepackage{amssymb}
\usepackage{bm}
\usepackage{epsfig}
\usepackage{graphicx}
\usepackage[colorlinks]{hyperref}
\usepackage[english]{babel}

\newcommand{\divv}{\mathop{\rm div}\nolimits}

\def\be{\begin{equation}}
\def\ee{\end{equation}}
\def\bea{\begin{eqnarray}}
\def\eea{\end{eqnarray}}

\begin{document}
\newcount\timehh  \newcount\timemm
\timehh=\time \divide\timehh by 60
\timemm=\time
\count255=\timehh\multiply\count255 by -60 \advance\timemm by \count255

\title{Phonon wind and drag of excitons in monolayer semiconductors}
\author{M.M. Glazov}

\affiliation{Ioffe Institute, 194021 St. Petersburg, Russia}

%\date{\today, \jobname.tex, printing time = \number\timehh\,:\,\ifnum\timemm<10 0\fi \number\timemm}

\begin{abstract}
We study theoretically the non-equilibrium exciton transport in monolayer transition metal dichalcogenides. We consider the situation where excitons interact with non-equilibrium phonons, e.g., under the conditions of localized excitation where a ``hot spot'' in formed. We develop the theory of the exciton drag by the phonons and analyze in detail the regimes of  diffusive propagation of phonons and ballistic propagation of phonons where the phonon wind is generated. We demonstrate that a halo-like spatial distribution of excitons akin observed in [Phys. Rev. Lett. {\bf 120}, 207401 (2018)] can result from the exciton drag by non-equilibrium phonons.
\end{abstract}

\maketitle

\section{Introduction}

Two-dimensional semiconductors based on transition metal dichalcogenides form a family of materials whose optical properties are dominated by excitons, electron-hole pairs bound by the Coulomb interaction~\cite{Mak:2010bh,Splendiani:2010a,Chernikov:2014a,RevModPhys.90.021001,Durnev_2018}. Sharp excitonic resonances with high oscillator strengths at the room temperature open up wide prospects for light-matter coupling studies and applications in the fields of photonics and polaritonics~\cite{Schneider:2018aa}.

Recently, an increased interest to the transport properties of excitons has appeared~\cite{Mouri:2014a,doi:10.1021/acs.jpclett.7b00885,Cadiz:2018aa,PhysRevLett.120.207401}. It is motivated by the possibility to access fundamental parameters of exciton dynamics such as momentum scattering time and diffusion coefficient as well as by the possibilities of applications of two-dimensional materials for the energy harvesting. In recent work~\cite{PhysRevLett.120.207401} different regimes of exciton transport 
have been revealed in monolayer WS$_2$: At low exciton densities the classical diffusion of excitons has been observed. An increase in the density of excitons enables the Auger process~\cite{abakumov_perel_yassievich,Sun:2014a,PhysRevB.89.125427,Mouri:2014a,Moody:16,Manca:2017aa,PhysRevX.8.031073} giving rise to an efficient non-radiative exciton-exciton annihilation. It changes the shape of the exciton distribution profile in the real space and gives rise to  an apparent increase of the observed diffusion coefficient. Finally, at even higher exciton densities the Gaussian-like distribution of the excitons evolves into long-lived halo shaped profiles with the dip in the middle. The effect is attributed to the memory effects resulting, e.g., from the heating of excitons due to the efficient Auger process~\cite{PhysRevLett.120.207401}.

The energy of overheated excitons can be dissipated by emission of phonons. It gives rise to a formation of a ``hot spot'' in the sample where the average kinetic energy of some of the  phonon modes exceeds by far the lattice temperature outside of spot. These nonequilibrium phonons start to propagate out of the hot spot bringing away extra energy and momentum. As a result of the exciton-phonon interaction the excitons tend to drift away from the excitation spot as well. In this scenario phonons can drag excitons just like non-equilibrium phonons push electrons in metals and semiconductors~\cite{legurevich:drag}, see Ref.~\cite{GUREVICH1989327} for review. Furthermore, if phonons propagate ballistically, the phonon ``wind'' is formed~\cite{keldysh_wind,kozub_drag,kozub_hot,PhysRevB.46.15058} resulting in unusual exciton propagation regimes somewhat resembling superfluid flow~\cite{PhysRevLett.69.2959,PhysRevB.46.15058,1996JETP...82.1180K,PSSB:PSSB45}. 

While the phonon drag of excitons has been studied in detail for bulk materials and conventional quasi-two-dimensional systems, its specific features in atomically-thin transition metal dichalcogenides have not been explored. In particular, while in bulk semiconductors and quantum wells the excitons mainly interact with three-dimensional phonons because a weak acoustic contrast is insufficient to strongly confine the phonons in the quantum well plane, in monolayer-thin materials and in van der Waals heterostructures the phonons are strongly confined in the two-dimensional layer. It results in the enhancement of the phonon density of states. This together with the fact that the exciton radii in these monolayer materials are quite small results in rather strong the exciton-phonon interaction in two-dimensional materials~\cite{PhysRevB.94.205423,PhysRevB.95.201202,Selig:2016aa,PhysRevLett.119.187402,shree2018exciton,Song:2013uq,PhysRevB.92.125431,PhysRevLett.122.217401}.

Here we develop the kinetic theory of the exciton transport in two-dimensional transition metal dichalcogenides in the case where excitons interact with non-equilibrium phonons. In Sec.~\ref{sec:kin} we formulate the problem and present the kinetic equations for the exciton and phonon distribution functions. Next, in Sec.~\ref{sec:diff} we reduce the kinetic equations to an effective drift-diffusion model for excitons. Section~\ref{sec:results} contains analytical and numerical results on exciton propagation under the conditions of phonon drag and phonon wind. Brief conclusion is presented in Sec.~\ref{sec:concl}

\section{Kinetic theory}\label{sec:kin}

\subsection{Model}\label{subsec:model}

\begin{figure}
\includegraphics[width=\linewidth]{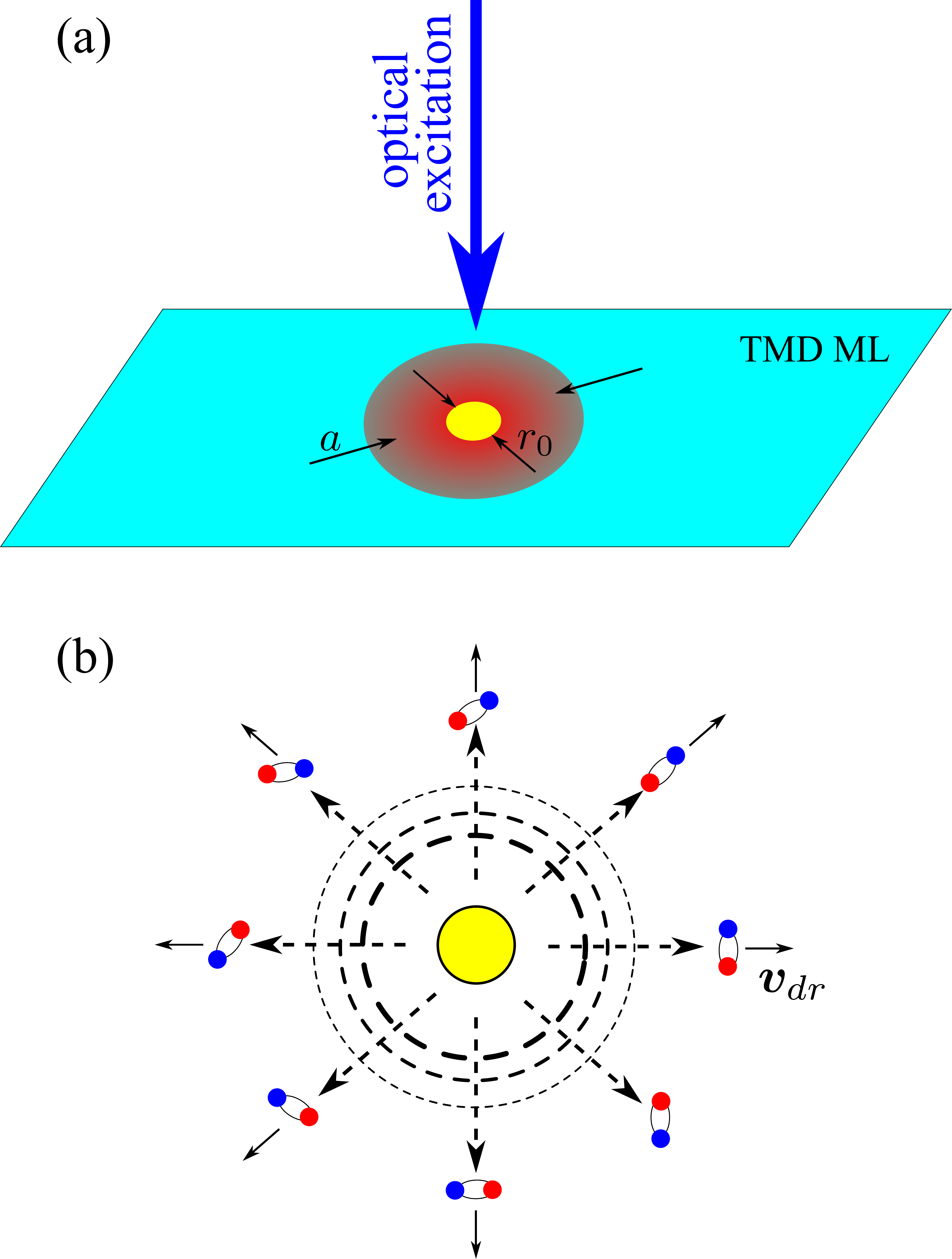}
\caption{(a) Schematics of the system and excitation. Red cloud shows excitons (effective radius $a$), yellow circle shows the phonon hot spot (effective radius $r_0$). (b) Scheme of phonon wind acting on excitons. Excitons are sketched as bound pairs of an electron (blue dot) and a hole (red dot); arrows illustrate exciton drift with the velocity $\bm v_{dr}$. }\label{fig:scheme}
\end{figure}

We consider a monolayer of transition metal dichalcogenide semiconductor and assume that at the initial time moment $t=0$ it is excited by a short and tightly-focused optical pulse with the photon energy exceeding the exciton resonance energy similarly to the experimental situation of Ref.~\cite{PhysRevLett.120.207401}. The absorption of the pulse results in the formation of {electron-hole pairs and excitons} which {lose} energy by phonon emission. We assume that initially the exciton density is high enough so that the Auger{-like exciton-exciton annihilation} process takes place as well. It results in the non-radiative ``bi-molecular'' recombination where one exciton recombines with its energy and momentum being transferred to another exciton which ends up in a highly-excited state~\cite{Takeshima1975,PhysRevB.54.16625,PhysRevB.73.245424,Sun:2014a,PhysRevLett.120.207401,Manca:2017aa,PhysRevX.8.031073}. It starts losing energy emitting more phonons. {Importantly, high energy phonons with almost flat dispersion that can include both optical phonons at the Brillouin zone center and the zone-edge modes serve as a major source of energy loss of non-equilibrium quasi-particles~\cite{gantmakher87}.} As a result, their velocity is close to zero and such phonons accumulate. They form a hot spot in monolayer semiconductor.  Eventually, owing to lattice anharmonicity optical phonons and zone edge acoustic phonons decay to the acoustic phonons with small wavevectors and almost linear dispersion. The latter propagate out of the  hot spot bringing away energy and momentum and produce the drag of the excitons. Another potential experimental option is to generate phonons by additional intense laser beam or electric current pulse, in which case one can control phonon distribution independently~\cite{moskalenko1,Moskalenko:1995aa,PhysRevLett.97.037401}. Schematically, the system under study is shown in Fig.~\ref{fig:scheme}(a) and the process of momentum transfer from non-equilibrium acoustic phonons with linear dispersion to excitons is shown in Fig.~\ref{fig:scheme}(b)

Thus, in order to describe the exciton dragging by non-equilibrium phonons it is sufficient to consider the acoustic phonons only taking into account the processes described above as a source of non-equilibrium acoustic phonon population. In what follows we describe the acoustic phonons by the distribution function $N_{\bm q}(\bm r,t)$, where $\bm q$ is the wavevector of the phonon, $\bm r$ is the position, and $t$ is time. Similarly, the excitons are described by the distribution function $f_{\bm k}(\bm r,t)$ with $\bm k$ and $\bm r$ being the exciton wavevector and its coordinate, respectively.  Below in Sec.~\ref{subsec:eqs} we present the set of coupled kinetic equations for the exciton and phonon distribution functions.

\subsection{Kinetic equations}\label{subsec:eqs}

The kinetic equation for the phonon distribution $N_{\bm q}(\bm r,t)$ describing phonon redistribution in the real and momentum space has the form
\begin{multline}
\label{phonon:kin}
{\partial N_{\bm q}\over \partial t} + {\bm V}_{\bm q}\cdot  \bm\nabla_{\bm r} N_{\bm q}  + \frac{N_{\bm q} - \bar N_{ q}}{\tau_p^{ph}}\\
 =-  \frac{N_{\bm q}}{\tau_0} + {Q_{ph-exc}\{N_{\bm q}\}}+  {\dot{N}_{\bm q}}(\bm r,t).
\end{multline}
Here ${\bm V}_{\bm q}=s\bm q/q$ is the phonon velocity in the state with the wavevector $\bm q$ with $s$ being the speed of sound, $\tau_0$ is the phonon lifetime, $\tau_p^{ph}$ is isotropization time of the phonon distribution function (phonon momentum relaxation time), and the overline denotes averaging over the directions of $\bm q$. The right hand side of Eq.~\eqref{phonon:kin} describes the processes where the total number of phonons changes, ${Q_{ph-exc}\{N_{\bm q}\}}$ is the phonon-exciton collision integral and ${\dot{N}_{\bm q}}(\bm r,t)$ is the generation rate in the course of the exciton cooling and Auger recombination. Here and in what follows we distinguish between the processes of acoustic phonon generation in the course of exciton energy relaxation via optical {and zone-edge phonons}, {described by the} term $\dot N$, and the interaction between the excitons and low-energy acoustic phonons, {described by the} term $Q_{ph-exc}$. {Equation~\eqref{phonon:kin} can be derived from the continuity equation for the distribution function taking into account the incoherent processes (phonon generation, annihilation and scattering) as collision integrals, whose form is specified below.}

The kinetic equation for the exciton distribution function has a form similar to Eq.~\eqref{phonon:kin}
\begin{multline}
\label{exciton:kin}
{\partial f_{\bm k}\over \partial t} + {\bm v}_{\bm k}\cdot  \bm\nabla_{\bm r} f_{\bm k}  + \frac{f_{\bm k} - \bar f_{ k}}{\tau_p} \\
= - \frac{f_{\bm k}}{\tau_d} +{Q_{exc-ph}\{f_{\bm k}\}}+  {\dot{f}_{\bm k}}(\bm r,t).
\end{multline}
Here ${\dot{f}_{\bm k}}(\bm r,t)$ is the exciton generation rate due to optical excitation, $\bm v_{\bm k} = \hbar \bm k/m$ is the exciton velocity, $m$ is its translational motion mass, $\tau_p$ and $\tau_d$ are the momentum relaxation and {population decay time (lifetime)} of exciton, respectively, ${Q_{exc-ph}\{f_{\bm k}\}}$ is the exciton-phonon collision integral.
{Being interested in the exciton transport in the presence of non-equilibrium phonons we} also disregard any nonlinear effects resulting from exciton-exciton interaction such as exciton-exciton scattering, energy renormalization as well as the Auger recombination, see Sec.~\ref{sec:results} for brief discussion.

Kinetic equations~\eqref{phonon:kin} and \eqref{exciton:kin} are coupled via the collision integrals ${Q_{ph-exc}\{N_{\bm q}\}}$ and ${Q_{exc-ph}\{f_{\bm k}\}}$. These integrals describe the processes of phonon emission and absorption due to exciton-phonon interaction. {Hereafter we assume that the excitons are non-degenerate, $f_{\bm k} \ll 1$ and t}he collision integral operator
${Q_{ph-exc}\{N_{\bm q}\}}$  acting on the phonon distribution function has a form
\begin{multline}
\label{Q:ph:exc}
{Q_{ph-exc}\{N_{\bm q}\}} \\
= \sum_{\bm k} \left[{ W^{em}_{\bm k} (1+N_{\bm q})f_{\bm k} 
%(1+f_{\bm k-\bm q})
%\right.\\
%\left. 
-  W^{abs}_{\bm k} N_{\bm q}f_{\bm k}} %(1+f_{\bm k+\bm q})
 \right], 
\end{multline}
where the rates of the phonon emission, $W^{em}_{\bm k}$, and absorption, $W^{abs}_{\bm k}$,  are given by the Fermi golden rule
\begin{subequations}
\label{rates}
\begin{align}
W^{em/abs}_{\bm k} = \frac{2\pi}{\hbar} |M_q|^2 \delta(E_{\bm k} - E_{\bm k - \bm q} \mp \hbar s q)
%,\\
%W^{abs}_{\bm k} = \frac{2\pi}{\hbar} |M_q|^2 \delta(E_{\bm k} - E_{\bm k + \bm q} + \hbar s q),
\end{align}
\end{subequations}
with $M_q$ being the matrix element of the exciton-phonon interaction~\cite{shree2018exciton} and $E_{\bm k} =  \hbar^2 k^2/2m$ is the exciton dispersion. For simplicity {and better clarity} we disregard here the complex band structure involving several valleys and also spin degrees of freedom: {Given similar masses and phonon coupling strengths across different valleys, it is further justified by inefficient intervalley and spin-flip scattering processes with low energy and momentum acoustic phonons}.

In what follows we assume that the phonon occupancies are high, $N_{\bm q} \gg 1$, in agreement with the situation we are describing. This condition is fulfilled since the energies of acoustic phonons $\hbar s q$ interacting with excitons is typically much smaller than exciton energy~\cite{gantmakher87,shree2018exciton}, thus, {under experimental conditions it is much smaller than the temperature both of the lattice and of the excitons. %In contrast, the exciton density is assumed to be sufficiently low to disregard nonlinear effects and assume $f_{\bm k} \ll 1$. 
In this regime it is sufficient to consider only stimulated processes and the collision integral can be further simplified as}
\begin{equation}
\label{Q:ph:exc:1}
{Q_{ph-exc}\{N_{\bm q}\}} = \mathcal W_{\bm q} N_{\bm q} ,
\end{equation}
where
\[
\mathcal W_{\bm q} = \frac{2\pi |M_q|^2}{\hbar}\sum_{\bm k}  \delta(E_{\bm k} - E_{\bm k - \bm q} - \hbar s q)(f_{\bm k}-f_{\bm k-\bm q}).
\]
The exciton interaction with acoustic phonons is just weakly inelastic, see Refs.~\cite{gantmakher87,shree2018exciton} for discussion. Thus, we have
\begin{equation}
\label{W:eff:1}
 \mathcal W_{\bm q} = \frac{2\pi |M_q|^2}{\hbar}\sum_{\bm k}  \delta(E_{\bm k} - E_{\bm k - \bm q})(\hbar s q \bar f_{k}'+ \delta f_{\bm k}- \delta f_{\bm k-\bm q}),
\end{equation}
where $\bar f_{k}' = d\bar f/dE_k$. Equations~\eqref{Q:ph:exc:1} and \eqref{W:eff:1} demonstrate that exciton-phonon collisions induce (generally anisotropic) correction to the phonon lifetime, it will be ignored in what follows.

Under the same assumptions as above (i.e., low exciton occupancy, $f_{\bm k} \ll 1$, and high phonon occupancy, $N_{\bm q} \gg 1$) we arrive at the following form of the exciton-phonon collision integral:
\begin{multline}
\label{Q:exc:ph}
{Q_{exc-ph}\{f_{\bm k}\}} \\
=\sum_{\bm q}  N_{\bm q} 
\left[f_{\bm k+\bm q} W^{em}_{\bm k+\bm q} + f_{\bm k-\bm q} W^{abs}_{\bm k - \bm q} - f_{\bm k} (W^{em}_{\bm k} +W^{abs}_{\bm k}) \right]
\\
=
\frac{2\pi}{\hbar} \sum_{\bm q} |M_q|^2  (f_{\bm k+\bm q} - f_{\bm k}) \left[ N_{\bm q}  \delta(E_{\bm k+\bm q} - E_{\bm k} -\hbar s q) 
\right. 
\\
\left. 
+  N_{-\bm q}  \delta(E_{\bm k+\bm q} - E_{\bm k} +\hbar s q) \right].
\end{multline}

In the elastic approximation, where $\hbar s q$ in the energy conservation law is omitted, all the effect of exciton-phonon collisions is reduced to the isotropization of the exciton distribution function. In what follows we assume that this contribution to $\tau_p^{-1}$ is already included in the momentum relaxation rate in the kinetic equation~\eqref{exciton:kin}.

Thus, in order to describe the exciton drag by the phonons we need to take into account non-elasticity in Eq.~\eqref{Q:exc:ph} in the lowest non-vanishing order. To that end we arrive at the collision integral in the simple form
\begin{multline}
\label{Q:exc:ph:inel}
{Q_{exc-ph}\{f_{\bm k}\}} \\
{= 
 2\pi s \bar f_{k}'\sum_{\bm q} q |M_q|^2  (N_{\bm q} - N_{-\bm q})  \delta(E_{\bm k+\bm q} - E_{\bm k})}.
\end{multline}
In derivation of Eq.~\eqref{Q:exc:ph:inel} we assumed that the anisotropic in the momentum space part of the exciton distribution function is much smaller that the isotropic one and neglected it. {We stress that the right hand side of the collision integral $Q_{exc-ph}\{f_{\bm k}\}$ Eq.~\eqref{Q:exc:ph:inel} is non-zero due to the fact that the phonons propagate out of the hot spot, Fig.~\ref{fig:scheme}. Correspondingly, the spatial gradient of $N_q$ produces the anisotropic part of the phonon distribution function $N_{\bm q} - N_{- \bm q} \ne 0$.}

In the following section we reduce the kinetic equations to the drift-diffusion equation for excitons which is suitable for description of the phonon drag.

\section{Exciton drift-diffusion model}\label{sec:diff}

\subsection{Phonon distribution function}\label{subsec:ph}

Hereafter we assume that propagating excitons provide minor effect on phonon distribution function. This is reasonable because the phonons are generated while the major part of the exciton population relaxes in energy and decays due to the Auger process. 
%There are few remaining excitons which diffuse or drift out of the excitation spot. 
The collision integral \eqref{Q:ph:exc:1} is proportional to the number of excitons and is small. Consequently, we neglect the $Q_{ph-exc}$ in the right hand side of Eq.~\eqref{phonon:kin}. Hence, the solution of this equation is expressed via the Green's function $\mathcal G_{\bm q}(\bm r,t)$ as
\begin{equation}
\label{greens:sol}
N_{\bm q}(\bm r,t) = \int d\bm r' \int dt' \: \mathcal G_{\bm q}(\bm r-\bm r', t-t') {\dot{N}_{\bm q}}(\bm r',t'),
\end{equation}
The Green's function Fourier-transform satisfies the equation
\begin{equation}
\label{greens:F}
\left[-\mathrm i \omega + \frac{1}{\tau}+ \mathrm i (\bm V_{\bm q} \cdot \bm Q) \right] \mathcal G_{\bm q} (\bm Q,\omega)= \frac{\bar{\mathcal G}_{ q} (\bm Q,\omega)}{\tau_p^{ph}} +1
\end{equation}
with $1/\tau = 1/\tau_0 + 1/\tau_p^{ph}$. Its solution reads
\begin{equation}
\label{greens:F:1}
 \mathcal G_{\bm q} (\bm Q,\omega)= \tau \frac{\bar{\mathcal G}_{ q} (\bm Q,\omega)/\tau_p^{ph} +1}{1-\mathrm i \omega\tau + i Ql \cos{\varphi}},
\end{equation}
where $l=s\tau$, $\varphi$ is the angle between $\bm Q$ and $\bm q$, and the angular-average Green's function can be readily expressed as
\[
 \bar{\mathcal G}_{\bm q} (\bm Q,\omega)= \frac{P_{Q,\omega}\tau}{1-P_{Q,\omega} \frac{\tau}{\tau_p^{ph}}}, \quad P_{Q,\omega} = \frac{1}{\sqrt{(1-\mathrm i \omega \tau)^2+Q^2l^2}}.
\]
This yields the following closed form expression for the Green's function:
\begin{multline}
\mathcal G_{\bm q} (\bm Q,\omega)= 
\\
\frac{\tau \sqrt{(1-\mathrm i \omega \tau)^2+Q^2l^2}}{(\sqrt{(1-\mathrm i \omega \tau)^2+Q^2l^2} - \tau/\tau_p^{ph})(1-\mathrm i \omega\tau + \mathrm i Ql \cos{\varphi})}.
\end{multline}

We assume that at $t=0$ the ${\cal N}_q$ phonons with the wavevector absolute value $q$ were generated in the small spot of the area $\pi r_0^2$ at $\bm r=0$, Fig.~\ref{fig:scheme}(a).
Accordingly, we take the generation rate ${\dot{N}_{\bm q}}(\bm r,t)$ in the form~\footnote{Generalization of the results to the large hot spot size is trivial and beyond the scope of the present paper. Basic results remain the same, see Appendix~\label{app:interplay}.} 
\begin{equation}
\label{generation:phonons}
{\dot{N}_{\bm q}}(\bm r,t)=\pi r_0^2 {\cal N}_q\delta(\bm r)\delta(t).
\end{equation} 
In this case we get:
\begin{multline}
N_{\bm q}(\bm r,t) = \pi r_0^2 {\cal N}_q  \mathcal G_{\bm q}(\bm r, t)
\\
 = \pi r_0^2 {\cal N}_q  \int \frac{d^2Q}{(2\pi)^2} \int \frac{d\omega}{2\pi} \mathcal G_{\bm q}(\bm Q,\omega) \text{e}^{i(\bm Q \cdot \bm r - \omega t)}.
\end{multline}

In what follows we consider two important limits for phonon propagation. First one is the case of \emph{ballistic propagation} where the momentum relaxation of phonons in unimportant, $\tau_0/\tau_p^{ph} \ll 1$. In this regime we have $\mathcal G_{\bm q}(\bm Q,\omega) \approx (\tau_0^{-1}-\mathrm i \omega + i Q s \cos{\varphi})^{-1}$, and
\begin{equation}
\label{N_ball}
N_{\bm q}^\text{ball}(\bm r,t) = \pi r_0^2 {\cal N}_q \delta \left(\bm r - st{\bm q\over q}\right) \text{e}^{- t/\tau_0}.
%\\
%=  {\cal N}_q\text{e}^{- t/\tau_0} {\delta \left(r - st\right) \over 2\pi r} \delta(\varphi_{\bm q}-\varphi_{\bm r}).
\end{equation}
Equation~\eqref{N_ball} clearly shows that the phonons propagate along straight lines from the origin to the point $\bm r$ with the fixed speed $s$ and $\bm q\parallel \bm r$.

Second important regime of phonon transport is the \emph{diffusive propagation}, where the momentum scattering time $\tau_p^{ph} \ll \tau_0$. In the diffusion regime there are three small parameters: $\omega\tau_p^{ph}, Ql, \tau_p^{ph}/\tau_0 \ll 1$ and the distribution of phonons is practically isotropic in the momentum space. Since we are interested in the transfer of phonon momentum to excitons we need to account for its small anisotropy in the first order. As a result we have for the phonons Green's function  
\begin{equation}
\label{ph:greens:diff:new}
\mathcal G_{\bm q}(\bm Q,\omega) \approx {1 - i \bm V_{\bm q}\cdot \bm Q \tau_p^{ph}\over \tau_0^{-1}-i\omega+{\mathcal D_{ph}} Q^2}, 
\end{equation} 
where we introduced the phonon diffusion coefficient 
\begin{equation}
\label{ph:diff}
\mathcal D_{ph}={l^2\over 2\tau}={s^2\tau_p^{ph}\over 2}.
\end{equation}
{We stress that Eqs.~\eqref{ph:greens:diff:new} and \eqref{ph:diff} are valid in the regime of slow, diffusive, dynamics of phonon distribution, i.e., on the frequency scales $\omega \ll 1/\tau_p^{ph}$ and on the wavevector scales $Q \ll 1/l$, while the product $\omega\tau_0$ can be arbitrary. That is why it is sufficient to take into account only the lowest non-vanishing terms in $\omega$ and $Q$ in the denominator and recover the diffusion pole in the form $(-i\omega + {\mathcal D_{ph}} Q^2)^{-1}$. In this regime the short time  dynamics of the phonon distribution function can be included by appropriate modification of the initial condition~\eqref{generation:phonons}. }
Finally, for the phonon distribution we have
\begin{equation}
\label{N_diff}
N_{\bm q}^\text{diff}(\bm r,t) = 
\left(1 - \tau_p^{ph} \bm V_{\bm q}\cdot  \bm\nabla_{\bm r} \right){\cal N}_q r_0^2 {\text{e}^{-r^2/4\mathcal D_{ph}t - t/\tau_0}\over 4\mathcal D_{ph}t}.
\end{equation}

It is possible to recast Eq.~\eqref{N_diff} in somewhat different notations which are convenient for what follows. To that end we represent the wavevector average phonon distribution as $\bar N_{\bm q}^\text{diff}(\bm r,t)= k_B T(\bm r,t)/(\hbar s q)$,
where $k_B$ is the Boltzmann constant and  $T(\bm r)$ is the local (i.e., coordinate dependent) temperature of the small energy acoustic phonon subsystem~\footnote{Such description is valid for thermalized phonons which form Planck distribution due to anharmonic processes. Also we assume that $k_B T\gg \hbar s r_0^{-1}$, where $r_0$ is the hot spot size.}.
Hence,
\begin{equation}
\label{N_diff_T}
N_{\bm q}^\text{diff}(\bm r,t) = 
\left(1 - \tau_p^{ph} \bm V_{\bm q}\cdot  \bm\nabla_{\bm r} \right){k_B T(\bm r,t) \over \hbar s q} ,
\end{equation}
where $T(\bm r)$ satisfies standard heat conduction equation
\begin{equation}
\label{T:prop}
\frac{\partial T(\bm r, t)}{\partial t} = \mathcal D_{ph} \Delta T(\bm r, t) - \frac{T(\bm r,t)}{\tau_0} {+ \dot T(\bm r, t),}
\end{equation}
with $\dot T(\bm r, t)$ being the energy generation rate in the units of $k_B^{-1}$.
Equations~\eqref{N_diff_T} and \eqref{T:prop} allow one to calculate phonon flux for an intense local heating of the sample. {We emphasize here that $T(\bm r)$ is not necessary equals to the local lattice temperature and may not represent the heating of the material in the conventional sense.}

\subsection{Drift-diffusion equation for excitons}\label{subsec:exc:dr}

Let us now analyze the impact of exciton-phonon interaction on the exciton distribution. To that end we note that the collision integral~\eqref{Q:exc:ph:inel} can be recast as
\begin{equation}
\label{force:coll}
{Q_{exc-ph}\{f_{\bm k}\}} = - \bar f'_{ k} (\bm v_{\bm k} \cdot \bm {\mathcal F}_{\bm k}),
\end{equation}
where the scalar product of the exciton velocity and the effective force can be presented in the form
\begin{multline}
\label{force:coll1}
(\bm v_{\bm k} \cdot \bm {\mathcal F}_{\bm k}) = 
- \frac{ms}{\pi\hbar^2}   \int\limits_0^{2\pi} d\varphi_{\bm q} q |M_q|^2  \Theta(2k-q)
\\  \times (N_{\bm q} - N_{-\bm q})
.
\end{multline}
Here the phonon wavevector absolute value in the integral and the angle $\varphi_{\bm q}$ are interrelated as $q = -2k\cos{(\varphi_{\bm q}-\varphi_{\bm k})}$ by virtue of the energy conservation law
\[
\delta(E_{\bm k+\bm q} - E_{\bm k}) = {2m\over \hbar^2 q} \delta[q+2k\cos{(\varphi_{\bm q}-\varphi_{\bm k})}].
%\varphi_{\bm q} = \varphi_{\bm k} \pm \arccos{(-q/2k)},
%\qquad \mbox{or} \quad q = -2k\cos{(\varphi_{\bm q}-\varphi_{\bm k})}.
\]

In what follows we assume that the excitons propagate diffusively in the sample, i.e., we are interested in the exciton dynamics on the time scales which exceed by far the exciton momentum scattering time $\tau_p$. Therefore, under assumption $\tau_d\gg\tau_p$, the kinetic equation~\eqref{exciton:kin} with the collision integral in the form of Eq.~\eqref{force:coll} can be replaced by the effective drift diffusion equations. In order to derive these equations we introduce the exciton density $n(\bm r, t)$ and the exciton flux density $\bm j(\bm r,t)$ according to
\begin{equation}
\label{nj}
n = g\int \frac{d\bm k}{(2\pi)^2} f_{\bm k}, \quad \bm j = g\int \frac{d\bm k}{(2\pi)^2} \bm v_{\bm k} f_{\bm k}. 
\end{equation}
Here $g$ is the spin and valley degeneracy factor which takes into account the band structure of the transition metal dichalcogenides monolayers (within simplest approximation {where the intervalley excitons are neglected,} $g=4$ or $2$ depending whether both bright and spin dark excitons or only one species are involved). Naturally, the integration in Eqs.~\eqref{nj} picks the zeroth and first angular harmonics of the exciton distribution which determine, respectively, the particle density and flux.

Integrating Eq.~\eqref{exciton:kin} over the wavevectors and taking into account the exciton-phonon interaction does not change the number of excitons, we arrive at the continuity equation for the exciton density
\begin{equation}
\label{cont:n}
\frac{\partial n}{\partial t} + \divv{\bm j} + \frac{n}{\tau_d}=\dot n(\bm r, t),
\end{equation}
where 
\begin{equation}
\dot n(\bm r, t) = g\int \frac{d\bm k}{(2\pi)^2} {\dot{f}_{\bm k}}(\bm r, t)
\end{equation}
is the exciton generation rate.
The equation for the flux density $\bm j$ can be derived in the same way by multiplying Eq.~\eqref{exciton:kin} by $\bm v_{\bm k}$ and integrating over $\bm k$. In order to simplify the resulting expressions we assume that $\tau_p$ is independent of exciton energy and that the excitons on average have a Boltzmann distribution
\begin{equation}
\bar f_k \propto \frac{n(\bm r,t)}{T_x(\bm r,t)} \exp{\left(-\frac{E_k}{T_x(\bm r,t)} \right)},
\end{equation}
with the effective exciton temperature $T_x(\bm r,t)$ which, in general, is not equal to the acoustic phonon subsystem temperature $T(\bm r,t)$ and depends on the position and time. Calculation of the thermalization rate is beyond the scope of the paper. It can be carried out following Refs.~\cite{ivchenko1988energy,0953-8984-8-13-008,Selig_2018}. As a result we have
\begin{equation}
\label{exc:flux}
\bm j = -D_x \bm\nabla_{\bm r} n - \eta n \bm\nabla_{\bm r} T_x  + \frac{\tau_p}{m} \bm F(\bm r,t) n.
\end{equation}
Here the first term is responsible for the exciton diffusion with
\begin{equation}
\label{D:exc}
D_x(\bm r,t)=\frac{k_B T_x(\bm r,t)\tau_p}{m},
\end{equation}
being the exciton diffusion coefficient, the second term accounts for the Seebeck effect~\cite{ashcroft1976solid,Askerov:1994aa}, see also Ref.~\cite{2019arXiv190602084P} for advanced calculations of the Seebeck coefficient for excitons, with 
\begin{equation}
\label{Seebeck}
\eta = \frac{k_B \tau_p}{m},
\end{equation}
 and the last term accounts for the force $\bm F(\bm r,t)$ acting from the phonons on the excitons (phonon wind and drag effects),
\begin{equation}
\label{force:coll:1}
\bm F(\bm r, t) = -\frac{g}{(2\pi)^2 mn}\int {d\bm k}E_k \bar f_{k}' \bm{\mathcal F}_{\bm k}(\bm r, t).
\end{equation}
Note that if $\bm{\mathcal F}_{\bm k}$ is wavevector independent then $\bm F(\bm r,t) = \bm{ \mathcal F}(\bm r,t)$.
The set of Eqs.~\eqref{cont:n} and \eqref{exc:flux} can be combined into the single drift-diffusion equation for the exciton density
\begin{equation}
\label{dr:diff}
\frac{\partial n}{\partial t} + \frac{n}{\tau_d}= \bm\nabla_{\bm r}\left( D_x\bm\nabla_{\bm r}  + \eta \bm\nabla_{\bm r} T_x - \frac{\tau_p}{m} \bm F \right)n + \dot n(\bm r, t).
\end{equation}
The first term in parentheses  describes the exciton diffusion and two remaining terms describe the drift due to the Seebeck effect and interaction with phonons, respectively. Equation~\eqref{dr:diff} can be recast in a more compact form using the explicit expressions for $D_x$ and $\eta$, Eqs.~\eqref{D:exc} and \eqref{Seebeck}:
\begin{equation}
\label{dr:diff:1}
\frac{\partial n}{\partial t} + \frac{n}{\tau_d}=\frac{\tau_p}{m}\bm\nabla_{\bm r}\left[ \bm\nabla_{\bm r}\left(k_B T_x n\right) -  \bm F n \right] + \dot n(\bm r, t).
\end{equation}
Equation~\eqref{dr:diff} should be also supplemented by a similar equation for the exciton temperature $T_x(\bm r, t)$. 

The model formulated above describes the exciton diffusion and drift due the Seebeck effect on overheated excitons and exciton-phonon interaction. In fact, similar description is valid if excitons are interacting with other particles or quasi-particles in the two-dimensional crystal. For example, the excitons can be dragged by the fluxes of electrons and holes, which can be generated in the sample, e.g., due to the hot exciton dissociation. In this case, $\bm F$ has a meaning of the momentum transfer rate from the electrons and holes to the excitons.
We also note that non-linear effects due to exciton-exciton interactions, such as Auger recombination and renormalization of the exciton energy can be straightforwardly introduced into Eq.~\eqref{dr:diff}. Similar drift-diffusion description of excitons can be also applicable provided that exciton-exciton collisions~\cite{Note:xx} are more efficient as compared with the exciton-phonon and exciton-impurity scattering by analogy with the hydrodynamical regime of electron transport, see Refs.~\cite{Gurzhi_1968,PhysRevLett.117.166601} and references therein.

\section{Results}\label{sec:results}

In this section we present the results of the exciton propagation modelling in the framework of the drift-diffusion model, Eq.~\eqref{dr:diff}. First, we determine the effective force $\bm F(\bm r, t)$ acting from the phonons on the excitons and, second, we present the results of analytical and numerical calculations.

\subsection{Phonon-induced driving forces}

The exciton-phonon coupling in transition metal dichalcogenide monolayers is mainly governed by the deformation potential interaction~\cite{PhysRevLett.119.187402,shree2018exciton}. {In the long wavelength limit relevant for our work, where the product $q a_B \ll 1$, with $a_B$ being the exciton Bohr radius, the phonon-induced deformation of the crystalline lattice is homogeneous on the scale of the exciton size. Thus, the energy shift of the exciton state is simply provided by the variation of the band gap energy and only longitudinal acoustic phonons contribute to the effect. Accordingly, w}e present the matrix element $M_q$ in Eqs.~\eqref{rates} and \eqref{Q:exc:ph:inel} as 
\begin{equation}
\label{Mq}
|M_q|^2 = \Xi q,
\end{equation}
{ where $\Xi$ is a constant related to the conduction band, $D_c$, and valence band, $D_v$, deformation potentials as~\cite{shree2018exciton}
\[
\Xi = {\frac{\hbar}{2\varrho s \mathcal S}} (D_c - D_v)^2,
\] 
with $\mathcal S$ being the normalization area and $\varrho$ being the mass density of the two-dimensional crystal.}

For diffusive phonon propagation where $\tau_p^{ph} \ll \tau_0$, Eq.~\eqref{N_diff_T} yields
\begin{equation}
N_{\bm q}^{\rm diff} - N_{-\bm q}^{\rm diff} = - {2\tau k_B \over \hbar q^2} \bm q \cdot \bm\nabla T.
\end{equation}
and by virtue of Eqs.~\eqref{force:coll}, \eqref{force:coll1} and \eqref{force:coll:1} we have 
\begin{equation}
\label{force:diff:ph}
\bm F_{\rm drag} = - \frac{2 s\tau k_B}{\hbar^2}\Xi \bm\nabla T =- \frac{\tau}{\tau_{x}} k_B \bm\nabla T.
\end{equation}
In the last equation we introduced the characteristic exciton-phonon scattering time as
\begin{equation}
\label{time}
{1\over \tau_{x}} =
\frac{2m^2 s}{\hbar^4}\Xi.% \equiv {2\pi\over \hbar}{m\over 2\pi \hbar^2} \Xi {2ms\over \hbar}.
\end{equation}
Equation~\eqref{force:diff:ph} has a clear physical sense: $-k_B \bm\nabla T$ is the momentum per unit of time carried by the phonons and the factor $\tau/\tau_{x}$ (note that in diffusive regime $\tau =\tau_p^{ph}$) gives the fraction of phonon momentum transferred to exciton in the course of exciton-phonon interaction. Such regime of the exciton drift where phonons are diffusive is denoted as the \emph{phonon drag} regime.

It is instructive to compare the phonon drag force in Eq.~\eqref{force:diff:ph} and the effective force due to the Seebeck effect caused by the exciton temperature gradient, second term in Eq.~\eqref{exc:flux}, see also second term in the brackets in Eq.~\eqref{dr:diff}. Under our assumptions the corresponding exciton Seebeck force is simply given by
\begin{equation}
\label{Seebeck:f}
\bm F_S = - k_B \bm\nabla T_x.
\end{equation}
Provided that the temperature gradients of excitons and phonons are the same, the ratio $|\bm F_{\rm drag}|/|\bm F_S| = \tau/\tau_x <1$, because total scattering rate of phonons is larger than their rate of collisions with excitons. Generally, the temperatures of excitons and phonons as well as their gradients can differ, particularly, in the situation where the phonons are generated by additional light pulse. The study of competition between the Seebeck effect and phonon drag is beyond the scope of this work, we just stress that both effects produce additive contributions to the exciton flux. 

We recall that according to Eqs.~\eqref{N_diff} and \eqref{N_diff_T} the temperature distribution under pulsed excitation has a Gaussian form and the phonon drag force in Eq.~\eqref{force:diff:ph} acquires a form
\begin{equation}
\label{force:diff:ph:2}
\bm F_{\rm drag} = \frac{\tau}{\tau_x} k_B \pi r_0^2 T_0{\text{e}^{-r^2/4\mathcal D_{ph}t - t/\tau_0}\over 8(\mathcal D_{ph}t)^2} {\bm r}.
\end{equation}
We recall that $T_0$ is the effective temperature in the hot spot and $r_0$ is the hot spot radius, see Fig.~\ref{fig:scheme}. It is assumed that $r_0 \ll r$, $\sqrt{\mathcal D_{ph} t}$ and $\sqrt{D_x t}$ at the relevant space and time scales.

For ballistic phonon propagation where $\tau_p^{ph} \gg \tau_0$, Eq.~\eqref{N_ball} yields
\begin{multline}
N_{\bm q}^{\rm ball} - N_{-\bm q}^{\rm ball} =  {\cal N}_q\text{e}^{- t/\tau_0} \\
\times {\delta \left(r - st\right) \over 2\pi r} 
\left[\delta(\varphi_{\bm q}-\varphi_{\bm r}) - \delta(\varphi_{\bm q}-\varphi_{\bm r}+\pi)\right].
\end{multline}
Further, to obtain simple analytical expressions, we again approximate $\mathcal N_q$ as $\mathcal N_q = k_B T_0/\hbar s q$ with $T_0$ being the effective temperature of the hot spot and $r_0$ being its radius, Fig.~\ref{fig:scheme}(a). As a result, we obtain
\begin{equation}
\label{force:ball:ph}
\bm F_{\rm wind} = \frac{ k_B T_0 r_0^2}{s \tau_x}  {\delta \left(r - st\right) \text{e}^{- r/s\tau_0} \over 2\pi r} \frac{\bm r}{r} .
\end{equation}
It is instructive to generalize Eq.~\eqref{force:ball:ph} to a situation where the phonons are generated in the hot spot during a finite time $\tau_\epsilon$. To that end we replace $\delta(t)$ by $\tau_\epsilon^{-1} \exp{(-t/\tau_\epsilon)}$ in Eq.~\eqref{generation:phonons}. After some algebra we obtain instead of Eq.~\eqref{force:ball:ph} the following expression for the force
\begin{equation}
\label{force:ball:ph:1}
\bm F_{\rm wind} = \frac{ k_B T_0 r_0^2}{2\pi r s^2 \tau_x \tau_\epsilon}  {\Theta \left(st - r\right) \exp{\left(\frac{r-st}{s\tau_\epsilon} -\frac{r}{s\tau_0}\right)}} \frac{\bm r}{r} .
\end{equation}
Equation~\eqref{force:ball:ph:1} demonstrates that, neglecting phonon decay ($\tau_0\to \infty$), the momentum flux by the ballistic phonons decays as inverse distance due to phonon propagation~\cite{keldysh_wind} as it follows from the momentum conservation, Fig.~\ref{fig:scheme}(b). Finite lifetime of phonons gives rise to additional $\propto \exp{(-r/s\tau_0)}$ decay with $s\tau_0$ being the mean distance which phonon propagates before it decays. The factor $\Theta \left(st - r\right)\exp{[({r-st})/(s\tau_\epsilon)]}$ accounts for the retardation while phonons arrive from the hot spot.  The regime of exciton drift {due to the interaction with} ballistic phonons is denoted hereafter as a \emph{phonon wind}.

\subsection{Solution of drift-diffusion equation}

We start with the simplified analysis of the exciton drift-diffusion equation~\eqref{dr:diff} and identify the most important limits. We also confirm our analytical results by the numerical solution of the drift-diffusion equation.

It is noteworthy that in the diffusive regime of exciton propagation the exciton momentum is lost at a very short time scale $\tau_p$. That is why as soon as the force field $\bm F$ vanishes, i.e., due to the phonon diffusion or finite phonon lifetime, the exciton distribution evolves according to the diffusion equation, spreads in the space and decays due to the finite exciton lifetime $\tau_d$. By contrast, while substantial number of phonons are present, a competition between the drift and diffusion takes place. The diffusive flux of excitons $\propto D_x |\bm\nabla n|\sim D_x n/a$, where $a$ is the exciton spot size, Fig.~\ref{fig:scheme}(a). The drift flux of excitons is given by $\tau_p Fn/m$. Thus, the drift of excitons dominates provided 
\begin{equation}
\label{drift}
\frac{\tau_p}{m} F \gg \frac{D_x}{a}, \quad\mbox{or}\quad F \gg \frac{k_B T_x}{a}.
\end{equation}
The latter condition follows from the exciton diffusion coefficient~\eqref{D:exc}. This condition can be naturally fulfilled in the phonon wind regime. {The situation can be different in the case of the Seebeck or the phonon drag effect. Indeed, taking into account that the force is given by the temperature gradient [see Eqs.~\eqref{force:diff:ph} and \eqref{Seebeck:f}] and making crude estimates with $\tau_p^{ph}\sim \tau_x$ we obtain that the drift and diffusion provide comparable contributions to the exciton dynamics if $T\lesssim T_x$.}

In order to obtain the analytical result we assume that the drift dominates, i.e., that Eq.~\eqref{drift} is fulfilled. In this case the diffusive term in Eq.~\eqref{dr:diff} can be neglected, we neglect also the Seebeck effect putting $\bm\nabla T_x$ to zero. The remaining equation can be solved by the method of characteristics. Namely, we find the exciton dynamics from the second Newton's law which at $t\gg \tau_p$ reads
\begin{equation}
\label{v:dr}
\frac{d\bm r}{dt} \equiv \bm v_{ dr} = \frac{\tau_p}{m} \bm F(\bm r, t).
\end{equation}
Here $v_{dr}$ is the exciton drift velocity in the force field $\bm F$.
The trajectories $\bm r(\bm \rho_0,t)$ where $\bm \rho_0$ is the initial position of the exciton at $t=0$ provide implicit dependence of $\bm \rho_0$ on $\bm r$ and $t$ which makes it possible to express the exciton distribution via the initial one. Assuming axial symmetry of the problem we obtain, see Appendix~\ref{app:analyt} for details
\begin{equation}
\label{nrt}
n(r,t) = n_0[\rho_0(r,t)]\frac{d\rho_0^2}{dr^2} e^{-t/\tau_d}.
\end{equation}

Let us first focus on the \emph{phonon wind regime} where the force field $\bm F(\bm r,t)$ is given by Eq.~\eqref{force:ball:ph:1}. It is noteworthy that exciton drift velocity, Eq.~\eqref{v:dr}, cannot exceed the speed of sound $s$. Indeed, the exciton-phonon scattering tends to equalize the velocities of the particles while the exciton momentum scattering processes result in the velocity dissipation. We assume in what follows that $v_{dr}\ll s$ in agreement with the assumptions made at derivation of Eq.~\eqref{dr:diff}; the effects where $v_{dr}$ is comparable with the speed of sound are beyond the scope of the present work, see, e.g., Ref.~\cite{PhysRevB.46.15058} where saturation effects were studied. Thus, in Eq.~\eqref{force:ball:ph:1} one can neglect $r$ as compared with $st$ and represent $\bm F_{\rm wind}$ in a very simple form
\begin{equation}
\label{Q:F}
\bm F_{\rm wind} = \frac{U(r,t)}{r} \frac{\bm r}{r},
\end{equation}
where the parameter $U(r,t) = U_0 \exp{(-t/\tau_\epsilon - r/s\tau_0)}$ with $U_0 \propto k_B T_0$ describes the hot spot efficiency. Equation~\eqref{Q:F} resembles the force field produced by a Coulomb center in a two-dimensional geometry. This is indeed consistent with the coordinate dependence of the momentum flux produced by the ballistic non-equilibrium phonons~\cite{keldysh_wind}. Equation~\eqref{v:dr} together with Eq.~\eqref{Q:F} can be easily integrated with the result
\begin{multline}
\label{implicit:wind}
e^{r/s\tau_0}(r-s\tau_0)-e^{\rho_0/s\tau_0}(\rho_0-s\tau_0)\\
=R_0(1- e^{-t/\tau_\epsilon}), \quad R_0 = \frac{\tau_p U_0}{m}\frac{\tau_\epsilon}{s\tau_0}.
\end{multline}
Particularly, for the excitons starting from the $\rho_0=0$ the dynamics is given by the following asymptotes depending on the relations between the parameters:
\begin{subequations}
\begin{align}
\label{wind:short}
&r(t) \approx \sqrt{2U_0t}, \quad t\ll \tau_0, \tau_\epsilon\\
&r(t) \approx \sqrt{2U_0\tau_\epsilon}, \quad t\gg \tau_\epsilon, \quad \tau_\epsilon \ll \frac{(s\tau_0)^2}{U_0},\label{wind:long:short}\\
&r(t) \approx s\tau_0, \quad t\gg \tau_0, \quad \tau_0 \ll \tau_\epsilon, \quad s\tau_0 \gg R_0,\label{wind:long:long}\\
&r(t) \approx s\tau_0 \ln\left(\frac{R_0}{s\tau_0}\right),~t\gg \tau_0,~\tau_0 \ll \tau_\epsilon,s\tau_0 \ll R_0.\label{wind:long:long:1}
\end{align}
\end{subequations}
Physically, at $t\to 0$ the excitons are strongly accelerated by the phonon wind since the force field~\eqref{Q:F} diverges as $1/r$ and giving rise to a formally infinite velocity in Eq.~\eqref{v:dr}. It results in $r\propto \sqrt{t}$ dependence, as demonstrated by Eq.~\eqref{wind:short}. As time goes by, the action of the phonon wind diminishes and excitons stop to be driven, $r=\mbox{const}$, either because the phonons vanish [short $\tau_\epsilon$, Eq.~\eqref{wind:long:short}], or because excitons eventually left the area covered by the phonon wind [long $\tau_\epsilon$, Eqs.~\eqref{wind:long:long} and~\eqref{wind:long:long:1}]. Thus, at long time scales the exciton dynamics is controlled by the diffusion and recombination processes.

\begin{figure}[t]
\includegraphics[width=0.9\linewidth]{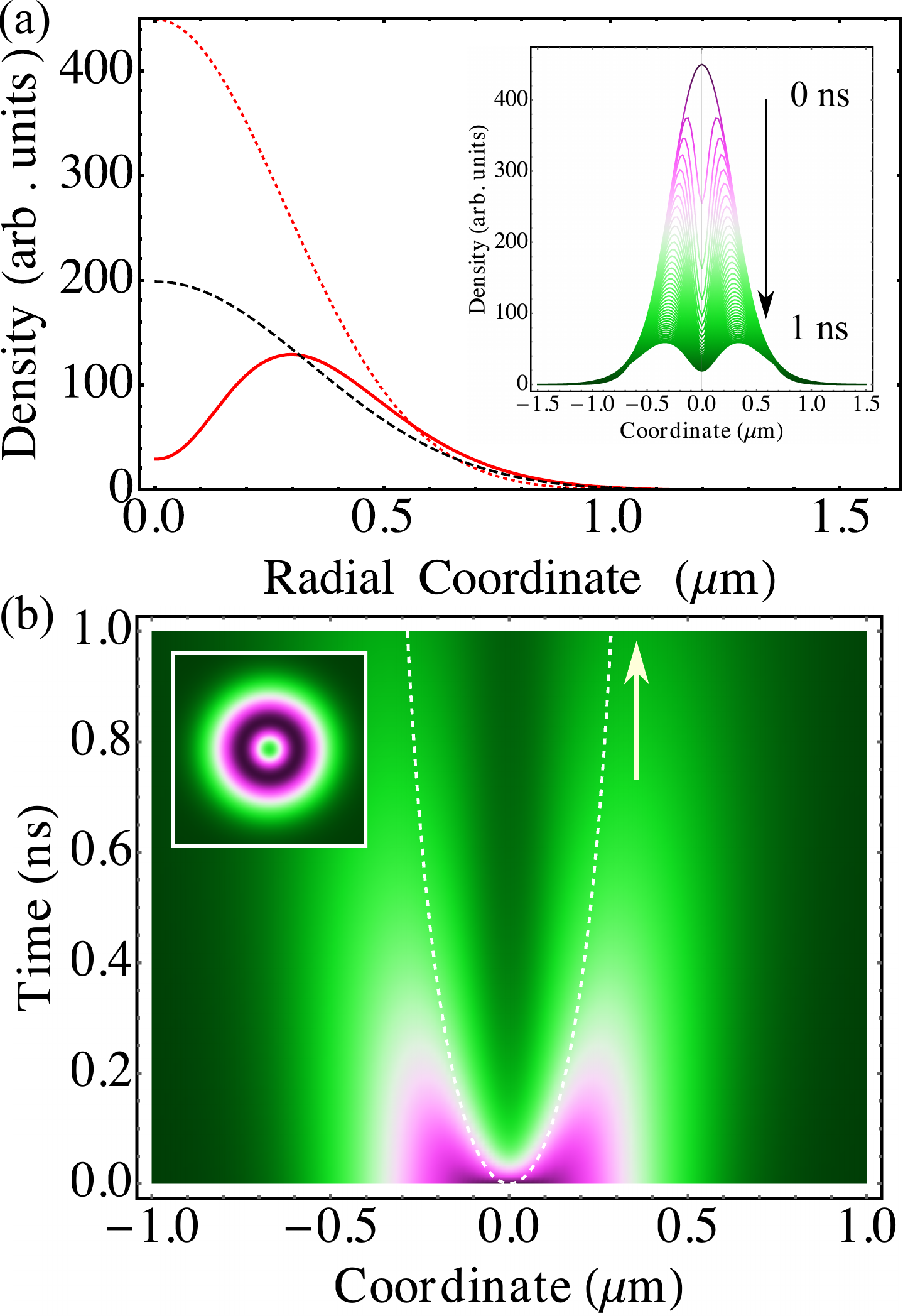}
\caption{Exciton propagation in the phonon wind regime. (a) Profiles of exciton density at $t=0$ (red dotted curve) and $t=0.5$~ns (red solid curve). Black dashed line shows the exciton profile at $t=1$~ns in the absence on non-equilibrium phonons ($U=0$). Inset shows profiles of the exciton density with time for $0\leqslant t\leqslant 1$~ns in equal steps. (b) False color plot of exciton density as a function of coordinate and time. White dashed lines show the solution of Eq.~\eqref{implicit:wind} at $\rho_0=0$. The arrow indicates the limiting value $r(t\to \infty)$  found from Eq.~\eqref{implicit:wind}. Inset shows the halo-like shape of the exciton cloud at $t=0.5$~ns. The parameters of exciton diffusion roughly correspond to Ref.~\cite{PhysRevLett.120.207401}: initial spot size $a=0.4$~$\mu$m, exciton lifetime $\tau_d=1.1$~ns, exciton diffusion coefficient $D_x=0.34$~cm$^2$/s; photon wind parameters are $\tau_\epsilon=\tau_d$, $s\tau_0=1$~$\mu$m, $R_0=0.0825$~$\mu$m. In our calculations we take large phonon lifetime of about $1$~ns which can be relevant to low temperatures and clean samples~\cite{Li:2013aa,Gu:2014aa} where the boundary scattering~\cite{2018arXiv181004467S} is unimportant. In the numerical calculation we also took into account the initial size of the phonon hot spot $r_0=0.08$~$\mu$m.
}\label{fig:wind}
\end{figure}

This  regime of exciton propagation is illustrated in Fig.~\ref{fig:wind}. {In our numerical calculations we took the initial distribution of excitons in the form
\begin{equation}
\label{init:n0}
n_0(\rho) = \frac{N_x}{\pi r_0^2} \exp{\left(- \frac{\rho^2}{a^2} \right)}
\end{equation}
Here $N_x = 2\pi \int_0^\infty n_0(\rho) \rho d\rho$ is the total number of excitons in the excitation spot and $a$ is the effective spot radius, $a^2 = 2\pi \int_0^\infty n_0(\rho) \rho^3 d\rho$ (half-width at half-maximum is $a\sqrt{\ln{2}}$). The exciton distribution was calculated by solving the drift-diffusion equation~\eqref{dr:diff} neglecting the Seebeck effect, i.e., putting $\eta=0$; the detailed analysis of the Seebeck effect on the exciton propagation is given in Ref.~\cite{2019arXiv190602084P}. We have also replaced the generation term $\dot{n}$ in the right-hand side with the initial condition~\eqref{init:n0} assuming instantaneous generation of excitons.  For the phonon wind mechanism we take the force field in the form of Eq.~\eqref{Q:F}.} 

{Figure~\ref{fig:wind}(a)} shows the  exciton distribution in the presence of wind (red solid line) and in the absence of wind (black dashed line) at a fixed time $t=1$~ns. The phonon wind markedly changes the distribution giving rise to the dip in the center which evolves in time as shown in the inset and also in more detail in Fig.~\ref{fig:wind}(b) where the false color plot of exciton distribution is shown. We also illustrate the propagation of excitons which started at the origin (dashed lines) calculated after Eq.~\eqref{implicit:wind}. It illustrates rapid expansion of the cloud and formation of a ring-shaped pattern [inset in Fig.~\ref{fig:wind}(b)] at short time scales, Eq.~\eqref{wind:short}, and saturation at longer times, Eq.~\eqref{wind:long:short}. The halo-like shape of the exciton cloud is a  result of the drift dominating over the exciton diffusion: the particles leave the hot spot faster than their distribution smoothens by the diffusion process. {The sensitivity of exciton density profiles to the parameters of the phonon system, $\tau_\epsilon$ and $\tau_0$, is briefly analyzed in Appendix~\ref{app:sens}. Taking shorter values of these times still results in the halo formation, but the size of the ring is respectively smaller, in agreement with approximate analytical expressions~\eqref{wind:long:short} and \eqref{wind:long:long}.}

\begin{figure}[t]
\includegraphics[width=0.9\linewidth]{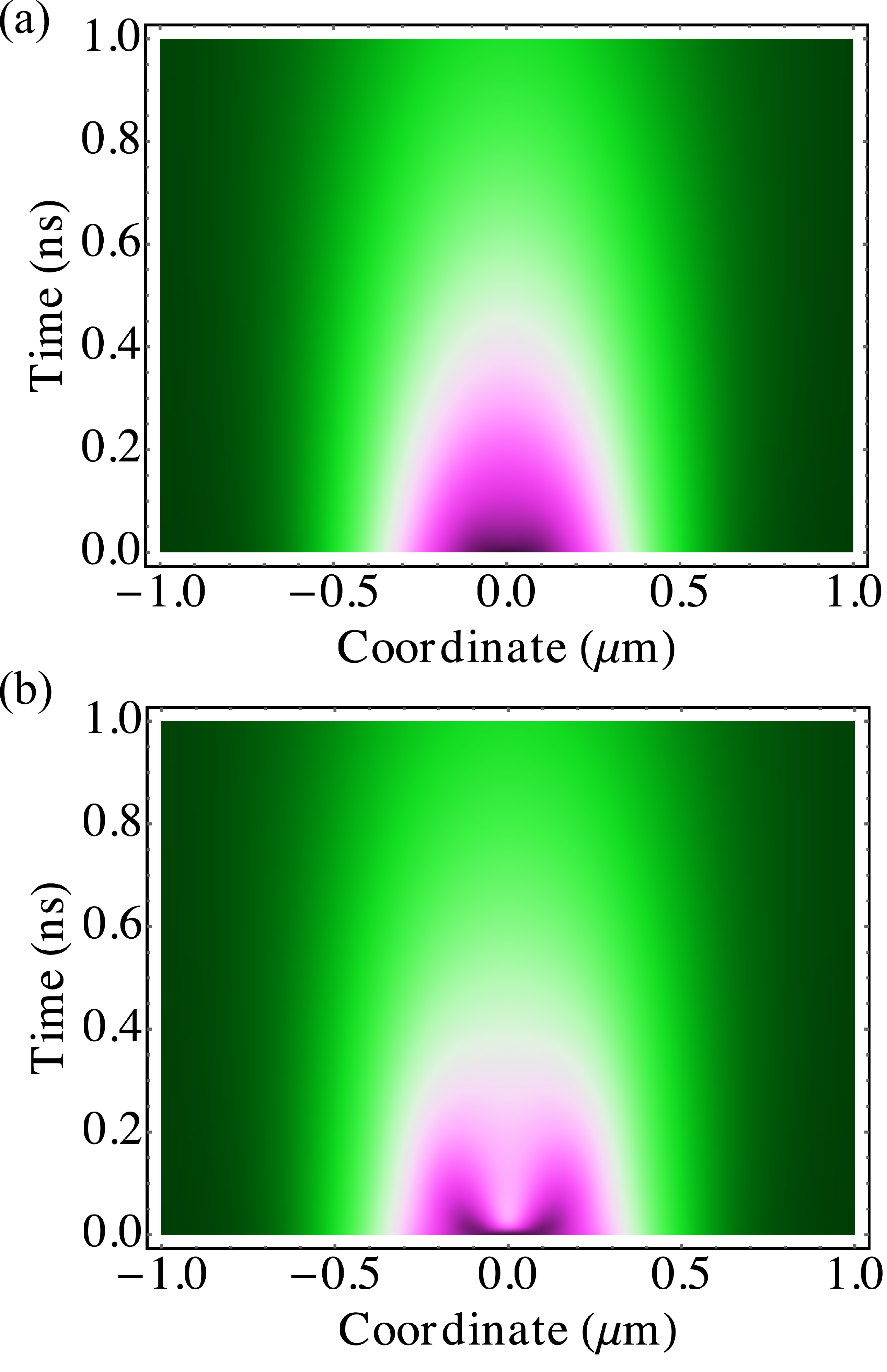}
\caption{Exciton propagation in the phonon drag regime. False color plots of exciton density for and effective temperature gradients (given in the units of drift velocity corresponding to the force at the phonon hot spot edge $r=r_0$ at $t=0$): (a) $F(r_0,0) \tau_p/m = 0.045$~$\mu$m/ns and (b) $F(r_0,0) \tau_p/m = 0.45$~$\mu$m/ns.  Other parameters are the same as in Fig.~\ref{fig:wind}: initial spot size $a=0.4$~$\mu$m, exciton lifetime $\tau_d=1.1$~ns, exciton diffusion coefficient $D_x=0.34$~cm$^2$/s; phonon lifetime $\tau_0=1.1$~ns, phonon hot spot radius $r_0=0.08$~$\mu$m and phonon diffusion coefficient $\mathcal D_{ph}=D_x$.
}\label{fig:drag}
\end{figure}

An interplay between the drift and diffusive behavior can be observed also in the \emph{phonon drag} regime as illustrated in Fig.~\ref{fig:drag}. {This calculation has been carried out similarly to the presented above, but the force field was taken in the form of Eq.~\eqref{force:diff:ph:2}.} Panels (a) and (b) demonstrate the exciton distribution evolution for two values of initial temperature gradient which correspond to relatively low and relatively high drift velocity of excitons, respectively. As expected, the significant drift of excitons and the halo formation are possible only provided that the drift velocity is large enough, so that the product $v_{dr} \tau_d\gtrsim \sqrt{D_x\tau_d}$. Similar behavior is  also possible in the regime of Seebeck effect, where the exciton drift is produced by the gradient of their temperature. The detailed study of the Seebeck effect requires self-consistent solution of the exciton drift-diffusion equation and the equation for the exciton temperature. This effect is beyond the scope of the present work.

Above we demonstrated the scenario for non-diffusive propagation of excitons: energy relaxation of excitons results in the formation of highly non-equilibrium acoustic phonons which, in turn, drag excitons out of the excitation spot. The spatial profile of the exciton density acquires a non-monotonic halo-like shape with the dip in the middle and increased exciton density on the periphery. Such profiles of the exciton density have been recently observed in Ref.~\cite{PhysRevLett.120.207401}. They were tentatively attributed to the memory effects such as the heating of the exciton gas and subsequent variation of the Auger recombination rate. The non-radiative exciton-exciton annihilation additionally decreases the number of excitons in the middle of the excitation spot. Here, the memory comes from the overheated phonon subsystem or from the elevated temperature of the excitons (in the case of the Seebeck effect induced drift). The full quantitative description of the experimental data requires also inclusion of the Auger recombination of excitons (which, according to our estimations, does not substantially qualitatively affect the profiles shown in Figs.~\ref{fig:wind} and \ref{fig:drag}) as well as the analysis of the phonon transport conditions. Particularly, at the room temperature the phonon lifetimes are expected to be sufficiently short ruling out the phonon wind effect. By contrast, at low temperatures the phonon wind could dominate the exciton drift. {In this work, we abstain from the detailed comparison of the predictions with experimental results of  Ref.~\cite{PhysRevLett.120.207401},  since a combination of several factors may affect the halo-like profile formation in the experiments at the room temperature: In particular, the combination of the wind and drag effects (on short- and long-timescales), possibly,  exciton-exciton or exciton-free carriers interaction, as well as the Seebeck effect (discussed as an origin of the halo-like profiles in Ref.~\cite{2019arXiv190602084P}) could result in the observed exciton density profiles.} Further experiments, e.g., aimed at studies of exciton propagation as a function of temperature would be helpful to finally establish the origin of the halo effect.

\section{Conclusion}\label{sec:concl}

We have developed analytical theory of exciton drift and diffusion in the presence of non-equilibrium phonons in two-dimensional transition metal dichalcogenides. We demonstrate that the flow of phonons can drag excitons out of the excitation spot giving rise to the halo-like shape of the exciton spatial profile. Different regimes of the drift are identified: for ballistic phonons the excitons are affected by the phonon wind, while for the diffusive phonons the exciton drift can be caused by the effective force resulting from the phonon temperature gradient.

\acknowledgements

The author is grateful to L.E. Golub, A. Chernikov, E. Malic, and S.G. Tikhodeev for valuable discussions. Partial support from Russian Science Foundation (project \# 19-12-00051) is acknowledged.

%\newpage

\appendix

\section{Analytical solution of the drift equation}\label{app:analyt}

Here we present analytical solution of Eq.~\eqref{dr:diff} at $D_x=0$ and $\tau_d \to \infty$. This drift equation for the density $n$ in the presence of the central force $\bm f$ has the following form:
\begin{equation}
\label{A:drift}
{\partial n \over \partial t} = - \bm \nabla (n \bm f), \qquad \bm f = {\bm r\over r} f(r,t).
\end{equation}
Then the drift equation~\eqref{A:drift} assumes the form
\begin{equation}
\label{A:drift:1}
{\partial n \over \partial t} =  -2 {\partial (rnf)\over \partial r^2}.
\end{equation}

We introduce the solution of the characteristic equation [second Newton's law, Eq.~\eqref{v:dr}] with the initial condition ${r(t=0)=\rho_0}$:
\begin{equation}
{d r\over dt} =  f(r,t), \qquad r=r(\rho_0,t), \qquad \rho_0=\rho_0(r,t).
\end{equation}
The following relation between the derivatives of $\rho_0^2$ takes place:
\begin{equation}
{\partial \rho_0^2 \over \partial t} =-2fr {\partial \rho_0^2 \over \partial r^2}.
\end{equation}

We now prove that the solution of Eq.~\eqref{A:drift:1} has the form of Eq.~\eqref{nrt} with $\tau_d\to\infty$:
\begin{equation}
\label{A:nrt}
n(r,t) = n_0[\rho_0(r,t)]{\partial \rho_0^2 \over \partial r^2}.
\end{equation}
Indeed the left hand side has the form
\begin{equation}
{\partial n \over \partial t} = n_0'{\partial \rho_0 \over \partial t}{\partial \rho_0^2 \over \partial r^2}
+ n_0 {\partial^2 \rho_0^2 \over \partial t\partial r^2},
\end{equation}
whereas the right hand side is:
\begin{equation}
-2 {\partial (rnf)\over \partial r^2} =  n_0' {\partial \rho_0 \over \partial r^2} {\partial \rho_0^2 \over \partial t}
+ n_0  {\partial^2 \rho_0^2 \over \partial r^2\partial t}.
\end{equation}
Evaluating the derivatives we see that the left and right hand sides coincide proving Eq.~\eqref{A:nrt}.

\begin{figure}[t]
\includegraphics[width=0.9\linewidth]{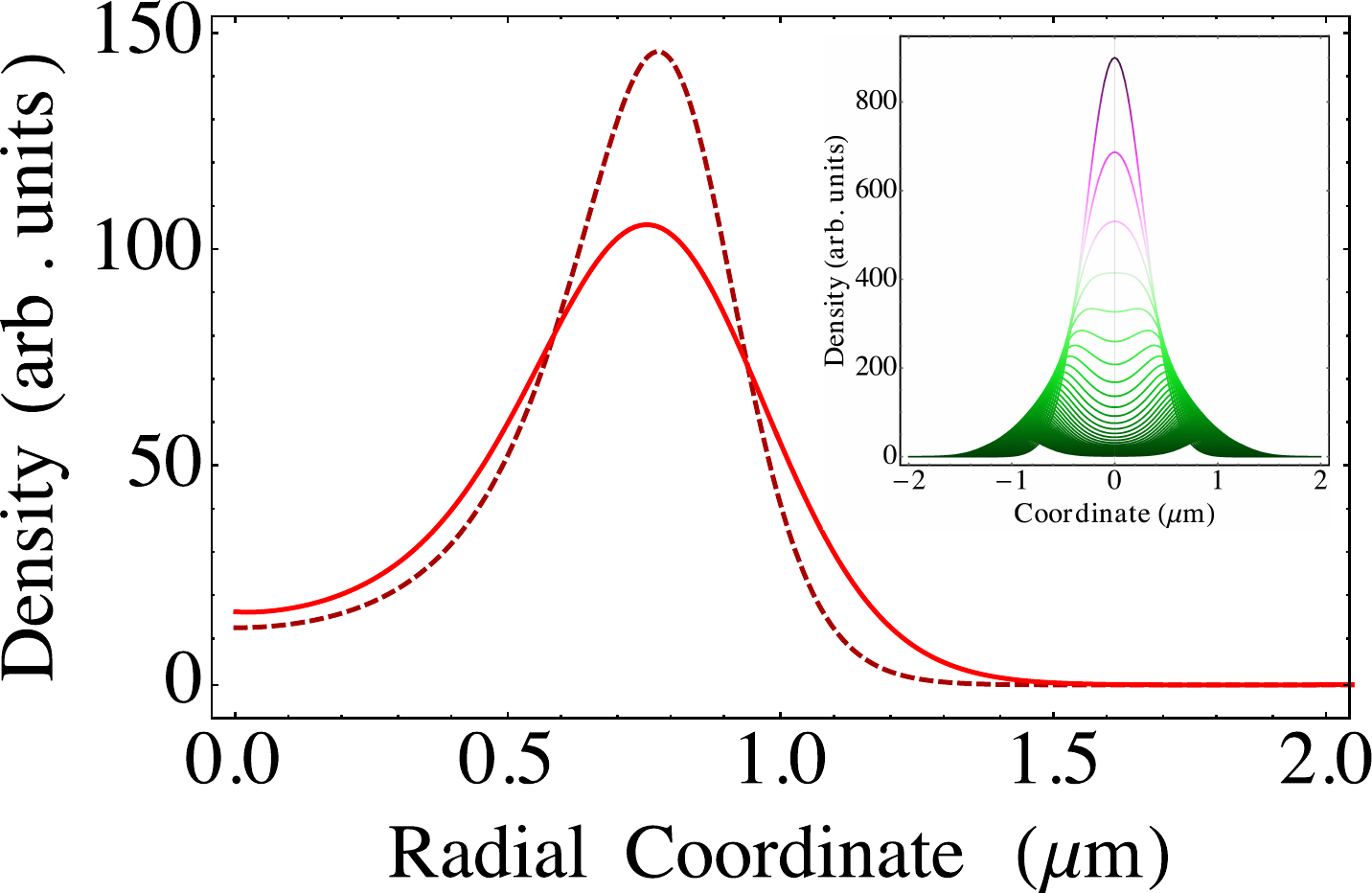}
\caption{Exciton propagation in the phonon wind regime. Comparison of the analytical solution, Eq.~\eqref{nrt}, with $\rho_0(r,t)$ given by Eq.~\eqref{char:rho:rho1} and the numerical solution of Eq.~\eqref{dr:diff} at $t=0.5$~ns. Inset shows profiles of the exciton density with time for $0\leqslant t\leqslant 1$~ns in equal steps. The parameters of exciton diffusion roughly correspond to Ref.~\cite{PhysRevLett.120.207401}: initial spot size $a={0.4}$~$\mu$m, exciton lifetime $\tau_d=1.1$~ns, exciton diffusion coefficient $D_x=0.34$~cm$^2$/s; photon wind parameters are $\tau_\epsilon=\tau_d$, $\sqrt{\tau_\epsilon\tau_p U_0/m}=0.9$~$\mu$m, phonon hot spot $r_0=0.4$~$\mu$m.
}\label{fig:wind:app}
\end{figure}

\section{Interplay of the drift and diffusion}\label{app:interplay}

It is instructive to provide the detailed comparison of the analytical (at $D_x=0$) and full numerical solutions of drift-diffusion equation~\eqref{dr:diff} in the regime of the phonon wind. We take in Eq.~\eqref{Q:F}
\[
U(r,t) = U_0 \left[1-\exp{(-r^2/r_0^2)}\right] e^{-t/\tau_\epsilon}.
\]

In order to produce the analytical solution we first find the exciton trajectories in the force field:
\begin{equation}
\label{char:rho:rho0}
\frac{\exp{[(r/r_0)^2]}-1}{\exp{[(\rho_0/r_0)^2]}-1} = \exp{\left[\frac{2 \tau_\epsilon \tau_p U_0}{m r_0^2}(1-e^{-t/\tau_\epsilon}) \right]}.
\end{equation}
Equations~\eqref{char:rho:rho0} or \eqref{char:rho:rho1} provide implicit dependence of $\rho_0$ on $r$ and $t$, i.e., the initial position of exciton to reach at the time $t$ the position $\rho$. Explicitly, it reads
\begin{equation}
\label{char:rho:rho1}
\rho_0(t,r) = r_0\sqrt{\ln{ \left[ 1+ \frac{\exp{(r^2/r_0^2)}-1}{ \exp{\left[\frac{2 \tau_\epsilon \tau_p U_0}{m r_0^2}(1-e^{-t/\tau_\epsilon}) \right]}} \right]}}.
\end{equation}
Figure~\ref{fig:wind:app} illustrates good agreement of the approximate analytical and exact numerical solution of the drift-diffusion equation.

\begin{figure}[t]
\includegraphics[width=0.9\linewidth]{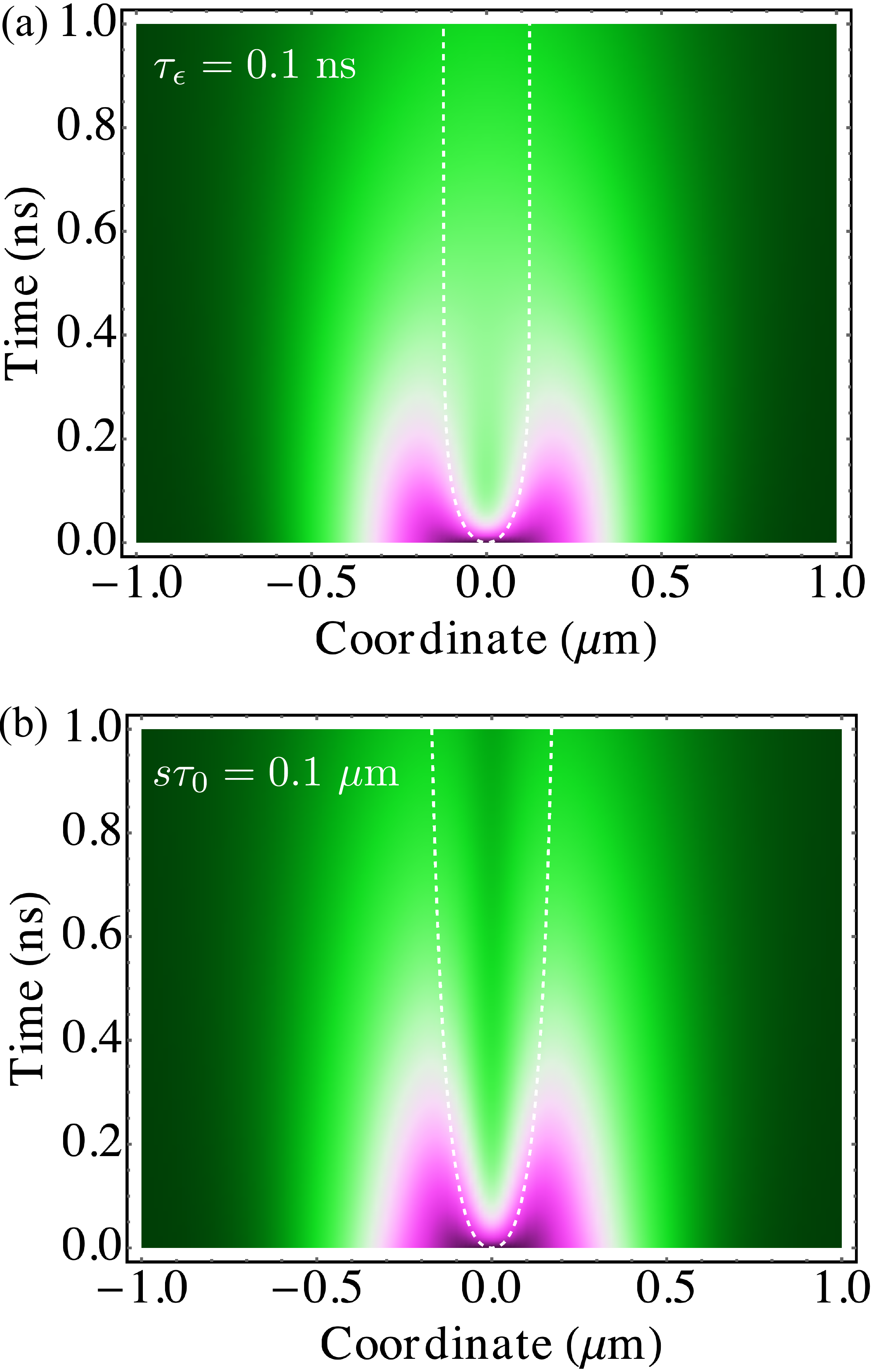}
\caption{{Exciton propagation in the phonon wind regime. (a) Calculated exciton density profile for the same parameters as in Fig.~\ref{fig:wind} but for $\tau_\epsilon=0.1$~ns. (b) Calculated exciton density profile for the same parameters as in Fig.~\ref{fig:wind} but for $s\tau_0=0.1$~$\mu$m.} 
}\label{fig:wind:app:1}
\end{figure}

\section{Halo sensitivity to the system parameters}\label{app:sens}

In Fig.~\ref{fig:wind:app:1} we analyze the sensitivity of the exciton density profiles in the phonon wind regime to the key parameters of the phonon system, which are largely unknown at this stage: hot spot generation time $\tau_\epsilon$ and phonon lifetime $\tau_0$. We took the same parameters of the system as we used to calculate Fig.~\ref{fig:wind} but used shorter value of $\tau_\epsilon$ in Fig.~\ref{fig:wind:app:1}(a) and shorter value of the product $s\tau_0$ in Fig.~\ref{fig:wind:app:1}(b). In both cases the halo formation is clearly seen, but the halo radius is smaller than that in Fig.~\ref{fig:wind}(b). This is consistent with approximate analytical expressions \eqref{wind:long:short} and \eqref{wind:long:long}, respectively. Minor differences from the analytical expressions results from the approximations used to derive Eqs.~\eqref{wind:short}, while the evolution of the halo peaks is well described by Eq.~\eqref{implicit:wind} (see dashed lines in Fig.~\ref{fig:wind:app:1}).

\newpage


\begin{thebibliography}{47}
\bibitem{Note:xx} The rate of exciton-exciton collisions can be estimated as $\sim n E_B a_B^2$, where $E_B$ and $a_B$ are the exciton binding energy and Bohr radius, respectively; the condensation-like effects can be ignored at $k_B T_x \gg \hbar^2 n/m$.
\end{thebibliography}
\end{document}